%% file: War_Stance_Detection.tex
\def\year{2021}\relax
\documentclass[letterpaper]{article} % DO NOT CHANGE THIS
\usepackage{aaai21}  % DO NOT CHANGE THIS
\usepackage{times}  % DO NOT CHANGE THIS
\usepackage{helvet} % DO NOT CHANGE THIS
\usepackage{courier}  % DO NOT CHANGE THIS
\usepackage[hyphens]{url}  % DO NOT CHANGE THIS
\usepackage{graphicx} % DO NOT CHANGE THIS
\usepackage{csquotes}
\usepackage{amssymb}
\usepackage{rotating}
\usepackage{amsmath}
\usepackage{natbib}  % DO NOT CHANGE THIS AND DO NOT ADD ANY OPTIONS TO IT
\usepackage{caption} % DO NOT CHANGE THIS AND DO NOT ADD ANY OPTIONS TO IT
\usepackage{array}
\usepackage{flushend}
\usepackage{setspace} 
\usepackage{makecell}
\usepackage[flushleft]{threeparttable}
\usepackage{booktabs,caption}
\usepackage[toc,page]{appendix}
\usepackage{subcaption}
\usepackage{rotating}
\usepackage{adjustbox} 
\usepackage{multirow}
\usepackage[T1]{fontenc}

\usepackage[usenames,dvipsnames]{xcolor}
\usepackage{tikz}
\usepackage{multirow}

\title{In the Eyes of the Bystander: Are the Stances on Different Conflicts Correlated?}
\author {
    Yiyao Tao\textsuperscript{*}, Hengyu Zhang\textsuperscript{*}, Babli Dey, Selenge Tulga, Hanjia Lyu, Jiebo Luo\textsuperscript{$\dagger$}\\
}
\affiliations {
    University of Rochester\\
    \textsuperscript{$\dagger$}jluo@cs.rochester.edu\\
    \textsuperscript{*} these authors contributed equally
}

\begin{document}

\maketitle

\begin{abstract}
Public opinion on international conflicts, such as the concurrent Russia-Ukraine and Israel-Palestine crises, often reflects a society's values, beliefs, and history. These simultaneous conflicts have sparked heated global online discussions, offering a unique opportunity to explore the dynamics of public opinion in multiple international crises. This study investigates how public opinions toward one conflict might influence or relate to another, a relatively unexplored area in contemporary research. Focusing on Chinese netizens, who represent a significant online population, this study examines their perspectives, which are increasingly influential in global discourse due to China's unique cultural and political landscape. The research finds a range of opinions, including neutral stances towards both conflicts and a statistical correlation between attitudes towards each, indicating interconnected or mutually influenced viewpoints. The study also highlights the significant role of news media, particularly in China, where state policies and global politics shape conflict portrayal, in impacting public opinion.
\end{abstract}

\section{Introduction}
Public opinion on international conflicts often mirrors the prevailing values, beliefs, and historical experiences of a society~\cite{feldman1988structure,dalton2013citizen}. By studying these opinions, we gain insights into what matters to people – such as peace, justice, human rights, and national sovereignty. This understanding is vital for comprehending the broader cultural and ethical contexts within which societies operate. Such significance cannot be overstated, especially in the context of the concurrent and highly discussed Russia-Ukraine and Israel-Palestine conflicts. These conflicts, unfolding almost simultaneously on the global stage, have not only captivated the attention of international audiences but have also sparked a multitude of discussions across various online platforms. This unique co-occurrence presents a prime opportunity to investigate the complexities and dynamics of public opinion in the face of multiple, simultaneous international crises.

One intriguing aspect of this scenario is how public opinions may evolve in response to the unfolding events of both conflicts. The interplay of opinions regarding two distinct, yet concurrently happening conflicts, offers a rich tapestry for analysis. Public sentiment towards each war may influence or mirror sentiments towards the other, revealing patterns in how global audiences perceive and react to international turmoil. This potential for intertwined perceptions is a relatively uncharted territory in contemporary studies. 

Furthermore, the current situation addresses a notable gap in existing research, which often focuses on understanding public opinion regarding a single conflict~\cite{imtiaz2022taking, chen2022public, kizilova2022assessing}. However, the simultaneous nature of the Russia-Ukraine and Israel-Palestine conflicts allows for a more holistic approach. It opens the door to exploring at a deeper level whether and how global public opinions shift, intertwine, or remain steadfast when faced with multiple, overlapping international crises.

The Russia-Ukraine and Israel-Palestine conflicts have also drawn intense discussions among Chinese netizens. China has one of the largest online populations in the world. The opinions and sentiments expressed by Chinese netizens can have a considerable impact on global discussions and narratives as China plays an increasingly bigger role in international affairs while at the same time maintaining at times significantly different cultures and values from the Western world. In this study, we seek to identify their favored sides and detect patterns in their preferences, thereby contributing to our understanding of public opinion dynamics. In particular, our research questions are:
\begin{itemize}
    \item \textbf{RQ1:} Which side do the Chinese netizens support in the Russia-Ukraine and Israel-Palestine conflicts, respectively?
    \item \textbf{RQ2:} Is there a correlation between the public's stance on one conflict and their stance on the other?
\end{itemize}

Our methodology adopts advanced data mining and machine learning techniques to analyze public stance. We extract a substantial volume of user comments from Weibo, which serves as a rich repository of public opinion on these conflicts. The data is then subjected to cleaning and preprocessing to ensure relevance and accuracy. To classify the stances, we employ a combination of sentiment analysis, logistic regression, multi-layer feedforward neural networks, and state-of-the-art pre-trained models like BERT and RoBERTa. This approach enables us to categorize the comments into distinct groups based on their stances towards the Russia-Ukraine and Israel-Palestine conflicts.

Moreover, our study incorporates statistical methods to assess the relationship between the stances on both conflicts. We also conduct a cross-analysis of these stances to better understand the intersection of opinions on these issues. Our approach provides an understanding of Chinese public opinion on these global conflicts, highlighting the complex nature of international opinion formation in the digital information age.

The study of these opinions is not just an academic exercise; it is a crucial endeavor in understanding the global psyche. Such insights can inform policymakers, diplomats, and international organizations, providing them with a nuanced and sophisticated understanding of public opinion. This, in turn, can aid in crafting more effective and responsive foreign policies, humanitarian efforts, and peacekeeping initiatives. In essence, the simultaneous unfolding of the Ukraine-Russia and Israel-Palestine conflicts provides a unique and timely laboratory for examining the complexities of public opinion in international affairs. By seizing this opportunity, we can deepen our understanding of how the public processes and responds to international conflicts, enriching both academic scholarship and practical policy-making in an increasingly interconnected world.

\section{Related Work}
Social network platforms have permeated all aspects of life. Weibo as a Chinese version of Twitter has attracted many netizens because of its free and convenient method of information interaction. According to \citet{huang2022stance}, netizens have a high degree of participation in public opinion events (likely higher than that of the United States), and the comments they release have significant sentiment tendencies. Additionally, Weibo serves as a platform for political expression and satire, allowing users to voice their opinions and criticisms, often using humor as a tool for commentary on political and social issues~\cite{guan2023moving}.

Despite stance detection on social media being a hard task, much research is being done to use various machine learning methods to model and train stance classifiers~\cite{kuccuk2020stance, imtiaz2022taking,lyu2022social, ahamed2022doctors, zhou2022fine}. Unlike sentiment analysis, \citet{aldayel2021stance} pointed out that stance detection mainly focuses on identifying a person’s standpoint or view toward an object of evaluation, either to be in favor of (supporting) or against (opposing) the topic. \citet{darwish2018predicting} proposed to use a sentiment analyzer to bootstrap the learning process of a supervised model when it comes to stance detection. The pre-trained models are a new approach to stance detection. \citet{ghosh2019stance} used the BERT model which generates representations of the words in the text through multiple transformer layers for stance detection. In our study, we conduct and compare transformer-based stance detection models and models relying on sentiment analysis to detect stance.

Several studies have used machine learning to obtain insights from the Russia-Ukraine or Israel-Palestine conflicts separately. \citet{chen2022public} collected data from the Chinese social media platform Weibo and employed the LDA model to determine the number of opinions on the Russia-Ukraine conflict. \citet{imtiaz2022taking} classified tweets into pro-Palestinian, pro-Israel, or neutral categories to understand the public stance on the Israel-Palestine conflict in 2021. However, there is limited research on a cross-analysis of both the Russia-Ukraine and Israel-Palestine conflicts. This study adopts a methodology akin to these studies but uniquely applies stance detection to both conflicts simultaneously, leveraging data from Weibo, one of the largest Chinese social media platforms.

\section{Material and Method}
\subsection{Data Collection}

In our study, we focus on the extraction of user comments from Weibo, comparable to Twitter in its influence. This platform is selected due to its extensive usage and the rich diversity of viewpoints it encompasses, particularly on global issues such as the Russia-Ukraine (RU) and Palestine-Israel (PI) conflicts. Our initial data scraping efforts are fruitful, amassing a substantial corpus of approximately 200,000 publicly available user comments. These comments, collected from various posts pertaining to the mentioned conflicts, offer a wide-ranging perspective on public opinion. A notable aspect of our data collection belongs to the temporal distribution of the comments. The majority of comments related to the RU conflict date back to 2022, with only about one-tenth including comments from 2023. In contrast, the comments on the PI conflict are specifically gathered from February to October 2023. Our methodological approach involves the strategic use of hashtags with keywords to efficiently filter relevant comments. For the PI conflict, the main keywords in hashtags included terms like ``Palestine'', ``Palestine-Israel Conflict'', ``Hamas'', ``Israel'', and ``Gaza Strip''. For the RU conflict, we use hashtags encompassing ``Russia'', ``Ukraine'', ``Russia-Ukraine Situation'', ``Russia-Ukraine Situation New Development'', and ``Russia-Ukraine War''. It is important to note that in Mandarin, searching for comments containing terms like ``Palestine'' may not capture those referencing the ``Palestine-Israel Conflict.''  This targeted approach enables the effective curation of comments that provide an understanding of the public's views on these complex and evolving international issues.

\begin{figure*}[ht!]
    \centering
    \includegraphics[width=\linewidth]{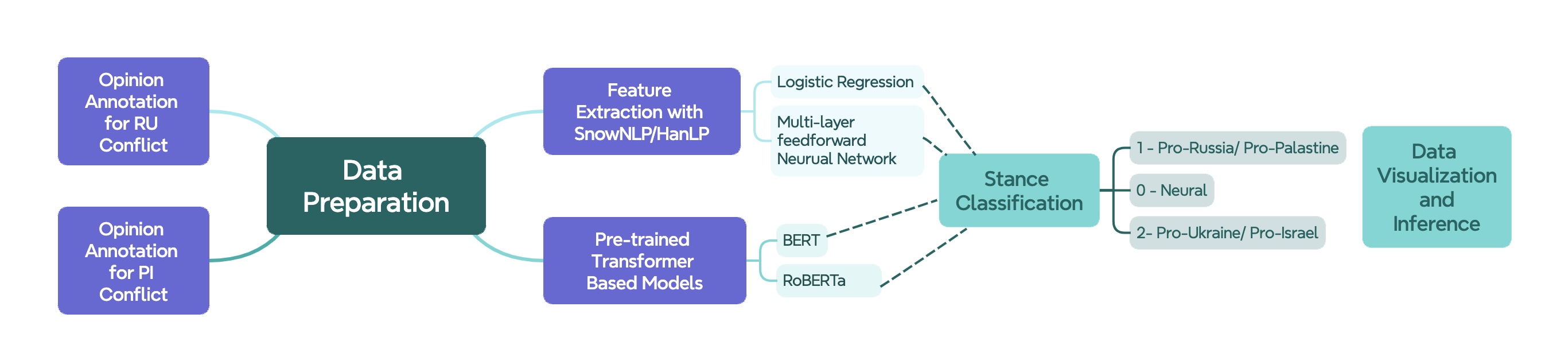}
    \caption{Stance detection process for analyzing opinions on both conflicts.}
    \label{fig:Figure 1}
\end{figure*}

\subsection{Data Cleaning and Preprocessing}

The raw data we collect is extensive, this requires a considerate cleaning process to ensure its relevance and accuracy. This crucial phase involves several specific steps: removing content within square brackets along with the brackets themselves, eliminating substrings beginning with ``source'', ``picture'', ``repost Weibo'', ``reply'', and retaining Chinese characters, and specified punctuations such as comma, period, and colon. Furthermore, entries with two or fewer characters are discarded to maintain data quality. The outcome of this cleaning is a significant reduction in data volume, leaving us with 4,022 valid, paired data entries.

A key aspect of our methodology is the division of data for model training. We filter the data based on the same user ID to achieve this, particularly focusing on users who commented on both the Russia-Ukraine and Palestine-Israel conflicts. Notably, most comments on Russia and Ukraine were from 2022, while those on Palestine and Israel were from 2023. This results in a marked decrease in the number of cross-users, a factor we account for in our data processing.

For model training, we treat the Palestine-Israel (PI) and Russia-Israel (RI) datasets separately to enhance the accuracy of our predictive results. Consequently, we develop two distinct models corresponding to these datasets. First, we annotate these 739 paired data entries ourselves. The labeling is based on the stances reflected in the comments: for both models, `0' indicates {\tt Neutral}. In the Russia-Ukraine model, `1' stands for {\tt Pro-Russian} and `2' for {\tt Pro-Ukraine}. Similarly, in the Palestine-Israel model, `1' is used for {\tt Pro-Palestine} and `2' for {\tt Pro-Israel}. The remainder of the dataset, comprising 3,283 paired data points, is then prepared for inference.

\subsection{Stance Detection}
Figure~\ref{fig:Figure 1} shows the pipeline of stance detection for two conflicts. In particular, Weibo provides a useful source of text for Chinese stance detection and opinion mining. We have conducted two methods to classify stances towards the ongoing Russia-Ukraine and Israel-Palestine conflicts. One is to combine sentiment analysis and other algorithms such as logistic regression and multi-layer feedforward neural network, the other approach is to use pre-trained transformer-based models including BERT~\cite{devlin2018bert} and RoBERTa~\cite{liu2019roberta}.

 The first approach applies pre-trained sentiment analysis tools such as SnowNLP and HanLP~\cite{he2021stem} to segment Chinese words and provide an array of features on the data. These features are put into a logistic regression/multi-layer feedforward neural network for text classification. The structure of the multi-layer feedforward neural network is described in Table~\ref{tab:performance_model}. Using the Russia-Ukraine dataset as an example, the model includes three hidden layers containing 512, 256, and 64 neurons, respectively. It also has an output layer with 3 neurons, corresponding to the categories: Pro-Russia, Pro-Ukraine, and Neutral. The model is implemented with a learning rate of 0.001 and a batch size of 32. It employs categorical cross-entropy as the loss function and Adam as the optimizer. Additionally, the training is conducted over 500 epochs.

\begin{table}[h]
\centering
% \small
\caption{Network structure.}
\label{tab:performance_model}
\begin{tabular}{ccccc}
\toprule[1.1pt]
Layer   & Activation Function & Output Shape\\
\midrule
Dense &  relu      &  [None, 512]  \\
Dense &  relu      &  [None, 256]  \\
Dense &  relu      &  [None, 64]   \\
Dense &  None      &  [None, 3]    \\
\bottomrule[1.1pt]
\end{tabular}
\end{table}

The second approach involves leveraging pre-trained transformer-based models to their fullest potential. We fine-tuned two BERT models ({\tt bert-base-chinese}) using our corpus, for the Russia-Ukraine and Israel-Palestine conflicts, respectively. Furthermore, we fine-tuned two RoBERTa models ({\tt chinese-roberta-wwm-ext-large}) for the same purpose. Both model types are configured to produce three distinct outputs (0, 1, 2) for identifying user stances. The training process spans 10 epochs, using a batch size of 16.	For training and testing models, we have manually annotated around 800 Weibo comments into three classes. The labeled data is then split into training and testing sets in the ratio of 4:1. These models are evaluated using separated testing sets, generating their performance metrics, including accuracy, precision, recall, and F1-score, which are detailed in Table~\ref{tab:model_performance}. Considering the overall model performance, we decide to employ RoBERTa for further study.

\begin{table*}[h]
\centering
\caption{Comparative performance of the models across the two conflicts. The best results are highlighted in \textbf{bold}.}
\label{tab:model_performance}
\adjustbox{max width=\linewidth}{
\begin{tabular}{lcccccccc}
\toprule[1.1pt]
\multirow{2}{*}{Model} & \multicolumn{4}{c}{Russia-Ukraine} & \multicolumn{4}{c}{Israel-Palestine} \\
 & Accuracy & Precision & Recall &   F1-Score& Accuracy & Precision & Recall & F1-Score \\ \midrule
SnowNLP with Logistic Regression & 0.60 & 0.60 & 0.60 & 0.60 & 0.66 & 0.65 & 0.66 & 0.60 \\
SnowNLP with Neural Network & 0.44 & 0.45 & 0.44 & 0.42 & 0.64 & 0.77 & 0.64 & 0.49 \\
HanLP with Logistic Regression & 0.41 & 0.41 & 0.41 & 0.38 & 0.61 & 0.66 & 0.61 & 0.49 \\
HanLP with Neural Network & 0.46 & 0.46 & 0.46 & 0.46 & 0.66 & 0.67 & 0.66 & 0.66 \\
BERT & 0.70 & 0.73 & 0.70 & 0.71 & \textbf{0.80} & 0.79 & \textbf{0.80} & \textbf{0.79} \\
RoBERTa & \textbf{0.74} & \textbf{0.75} & \textbf{0.74} & \textbf{0.74} & 0.73 & \textbf{0.84} & 0.73 & 0.76 \\ \bottomrule[1.1pt]
\end{tabular}
}
\end{table*}

\subsection{Data Analysis}

The first step involves calculating the percentage of each group's stance towards the Russia-Ukraine and Palestine-Israel conflicts, respectively. This is essential to understand the distribution of opinions among the users for each conflict. More specifically, to statistically assess whether the stances towards the Russia-Ukraine conflict and the Palestine-Israel conflict are related or independent, we apply the Chi-Squared test. This test would help determine if there is a significant association between the stances on the two conflicts. Further, to analyze the data by combining the stances for both conflicts, we create 9 pairs of groups (\textit{e.g.}, RU0-PI0, RU0-PI1, RU1-PI2, \textit{etc}.). This allows us to explore all the intersections of opinions on both issues. After pairing, we calculate the percentage of users in each of these combined groups to understand the prevalence of specific combined stances.

\section{Results}
\subsection{Between Conflicts}

\begin{figure}[h]
    \centering
    \includegraphics[width=\linewidth]{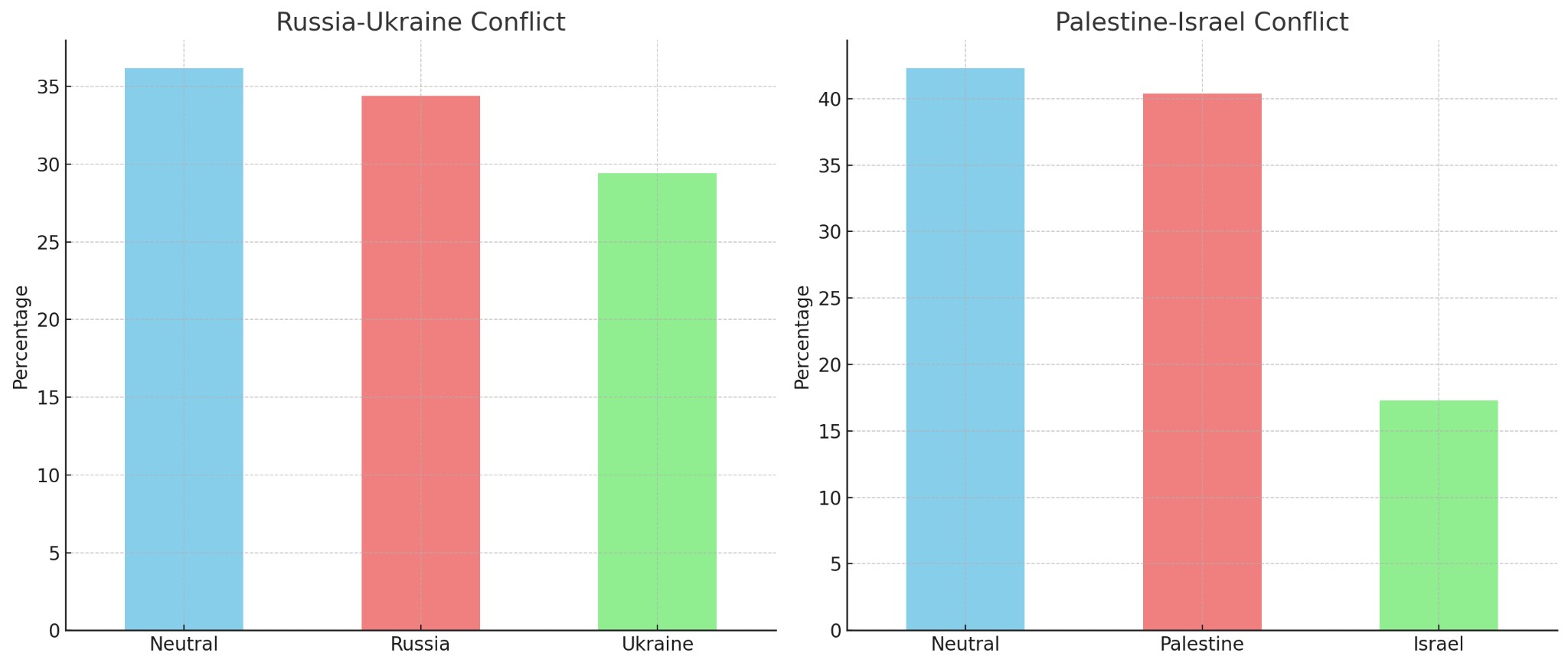}
    \caption{Distributions of users by the opinions towards the Russia-Ukraine and Palestine-Israel conflicts.}
    \label{fig:Figure 2}
\end{figure}

As shown in Figure~\ref{fig:Figure 2}, for the Russia-Ukraine conflict, among all users, 36.19\% hold a neutral stance, 34.39\% support Russia, and 29.42\% favor Ukraine. The distribution of stances is relatively even, with a slight majority being neutral. The difference between support for Russia and Ukraine is minor. On the other hand, regarding the Palestine-Israel conflict, 42.31\% of users in our study maintain a neutral stance. Of the remaining participants, 40.39\% support Palestine, while only 17.30\% favor Israel. This indicates a higher tendency towards neutrality, and the number of Palestine supporters is more than double that of Israel supporters.

These proportions suggest varied user stances and perceptions towards two conflicts. The relatively higher neutrality in both conflicts could indicate a more complex and uncertain view from the users' perspectives. The Russia-Ukraine conflict shows a balanced division of opinions. While the Palestine-Israel side has a more polarizing situation.

To investigate whether there is a statistical relationship between the stances on the two conflicts, we construct a contingency table. Table~\ref{tab:contingency} cross-tabulates the counts of users' stances for each conflict, providing a detailed view of how stances on one conflict relate to stances on the other.

\begin{table}[h]
\centering
% \small
\caption{Cross-analysis of Chinese netizens' stances on the two conflicts.}
\label{tab:contingency}
\begin{tabular}{ccccc}
\toprule[1.1pt]
RU \ PI   & Neutral (0) & Palestine (1) & Israel (2)\\
\midrule
Neutral (0)  &  636  &  375  &  177  \\
Russia  (1)  &  415  &  538  &  176  \\  
Ukraine (2)  &  338  &  413  &  215  \\
\bottomrule[1.1pt]
\end{tabular}
\end{table}

We then conduct a Chi-squared test for independence using this contingency table. We discover that there is a statistically significant association between the stances on the two conflicts ($p<.05$). The analysis reveals that the Chinese users' stances on the Russia-Ukraine conflict were \textit{not} independent of their stances on the Palestine-Israel conflict. 

\subsection{Mixture of Stances}

We will now examine the distribution of paired groups, which represent specific stances on both the Russia-Ukraine and Palestine-Israel conflicts. The percentages for each combined group are detailed in Table~\ref{tab:sentiment_analysis}.

\begin{table}[ht]
\centering
\caption{Stances analysis results.}
\label{tab:sentiment_analysis}
\begin{tabular}{ll}
\toprule[1.1pt]
Category & Percentage \\ \midrule
RU0-PI0 (Neutral on both) & 19.37\% \\
RU0-PI1 (Neutral on RU, Pro-Palestine) & 11.42\% \\
RU0-PI2 (Neutral on RU, Pro-Israel) & 5.39\% \\
RU1-PI0 (Pro-Russia, Neutral on PI) & 12.64\% \\
RU1-PI1 (Pro-Russia, Pro-Palestine) & 16.39\% \\
RU1-PI2 (Pro-Russia, Pro-Israel) & 5.36\% \\
RU2-PI0 (Pro-Ukraine, Neutral on PI) & 10.30\% \\
RU2-PI1 (Pro-Ukraine, Pro-Palestine) & 12.58\% \\
RU2-PI2 (Pro-Ukraine, Pro-Israel) & 6.55\% \\ \bottomrule[1.1pt]
\end{tabular}
\end{table}

\begin{figure}[h]
    \centering
    \includegraphics[width=\linewidth]{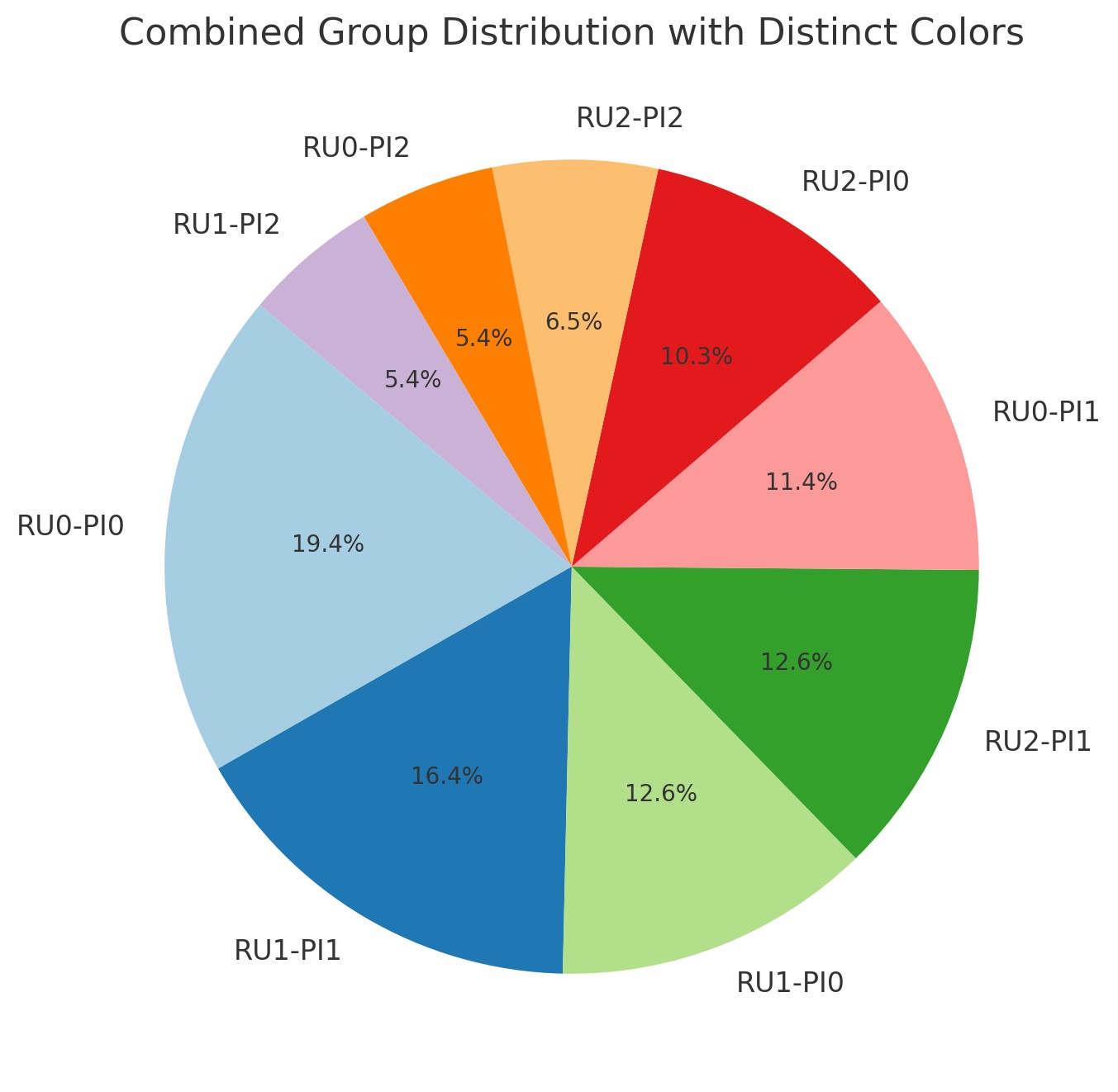}
    \caption{Percentage distribution of mixed stances towards the two conflicts.}
    \label{fig:Figure 3}
\end{figure}

\begin{figure}[ht]
    \centering
    \includegraphics[width=\linewidth]{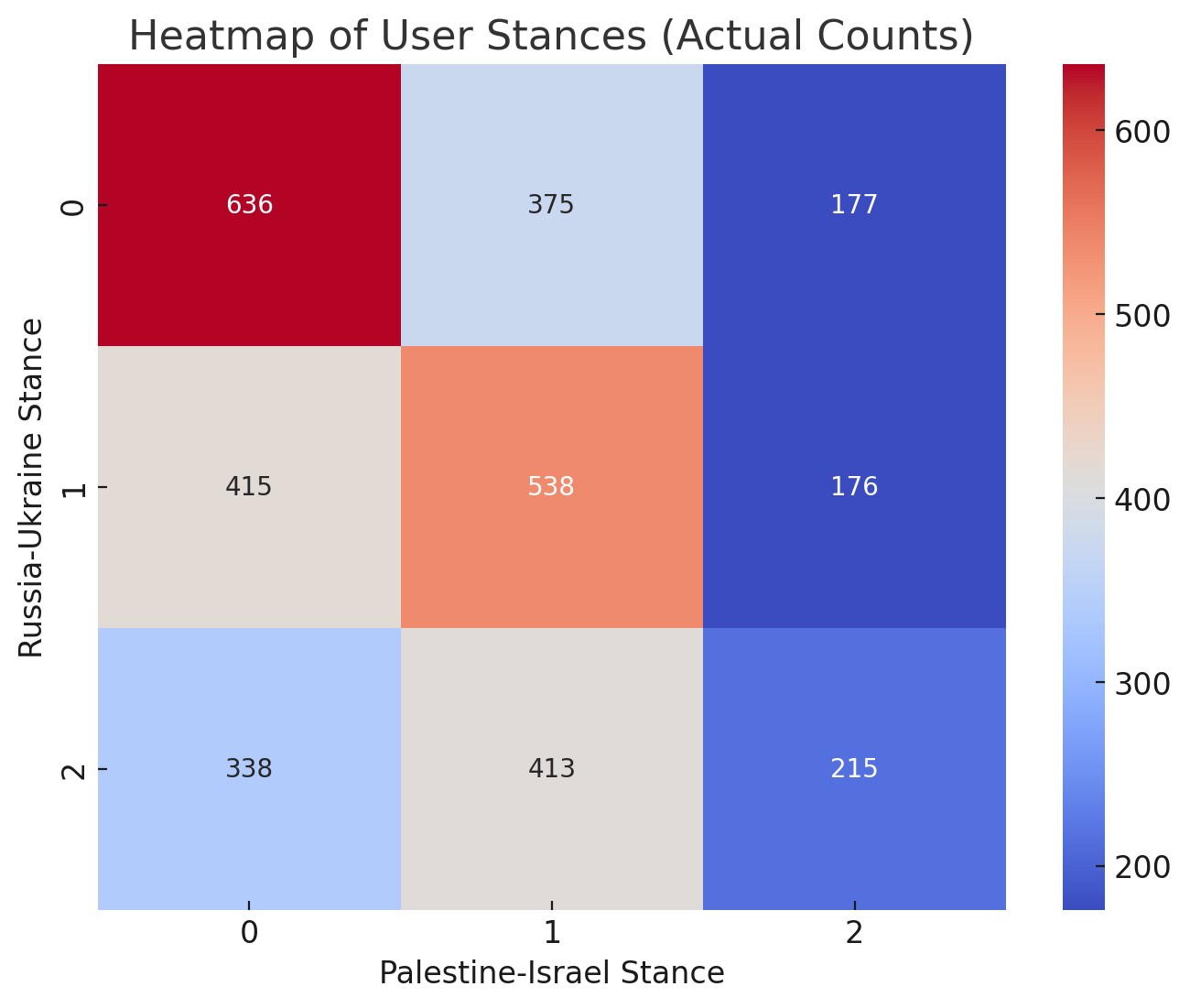}
    \caption{Heatmap of user stances towards the two conflicts.}
    \label{fig:Figure 5}
\end{figure}

The largest group is neutral towards both conflicts (RU0-PI0), suggesting a significant portion of users prefer a peaceful or neutral stance on these issues. The next largest groups are those with consistent stances towards one conflict but neutral towards the other (RU1-PI0, RU2-PI1, RU0-PI1). Groups with aligned stances towards both conflicts (\textit{e.g.}, RU1-PI1, RU2-PI2) represent a smaller but notable proportion. This heatmap shows the actual count of users in each combination of stances towards the Russia-Ukraine (Y-axis) and Palestine-Israel (X-axis) conflicts (Figure~\ref{fig:Figure 5}). Each cell represents the number of users with that specific combination of stances.

Next, we put effort into generating word clouds to help us understand the perception of each side's supporters, reasoning their choices, and finding discernible patterns. The inferences drawn from frequent words can be regarded as public opinion, formed at the individual level based on diverse online information sources and individual cognition.

\begin{figure}[h]
    \centering
    \includegraphics[width=\linewidth]{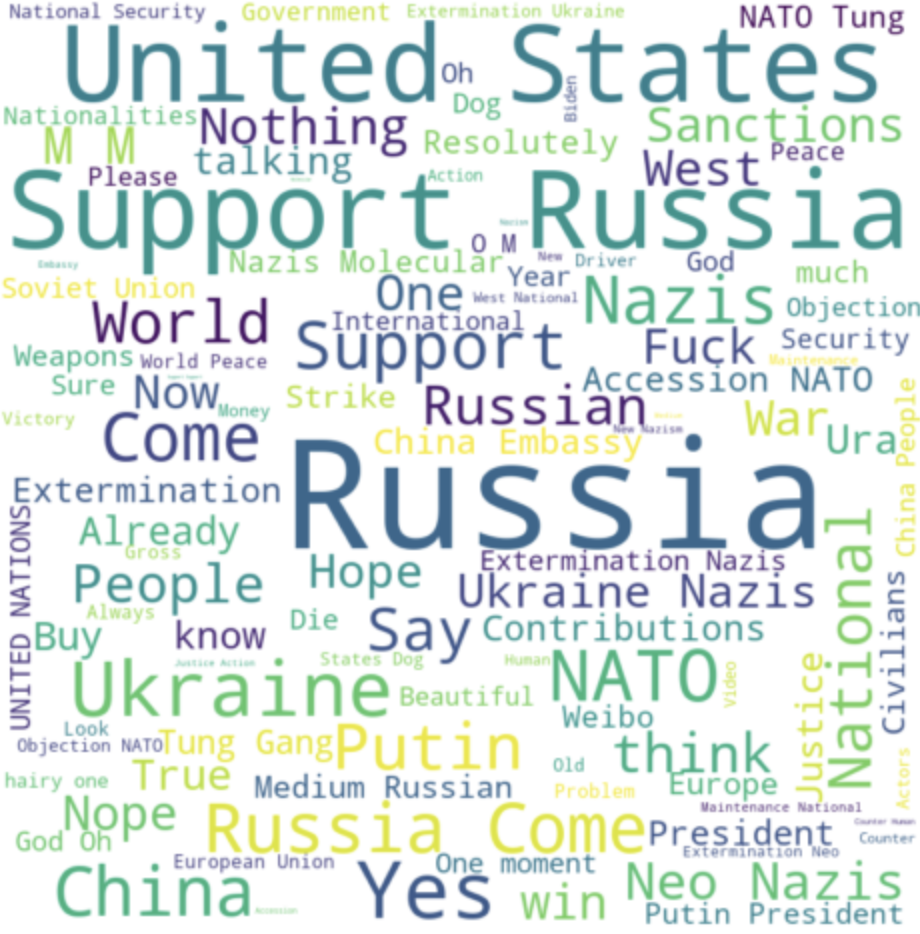}
    \caption{Tag cloud for Pro-Russia Weibo posts.}
    \label{fig:Figure 7}
\end{figure}

Given that China is the primary source of the data, the findings offer several intriguing insights. According to Figure~\ref{fig:Figure 7}, \textit{Support Russia}, \textit{Nazi}, \textit{NATO}, \textit{US}, \textit{Sanction}, and so on are frequent words among people supporting Russia, on querying these frequent words on Weibo, several reasons emerge that explain the strong support for Russia among these users. These reasons appear to be widely accepted or considered common sense by these netizens.  For instance, a common viewpoint blames NATO (North Atlantic Treaty Organization) for the conflict, asserting that NATO, led by the US, is engaging in an undeclared war against Russia to achieve its objectives. Additionally, there is noticeable dissatisfaction among Russia's supporters regarding the sanctions imposed by Western countries.  What's more, they believe in the coverage provided by news media~\cite{moscowtimes2022} to claim Ukraine as a Nazi, and the purpose of this conflict initialized by Russia is to cleanse Ukraine of Nazis.

\begin{figure}[h]
    \centering
    \includegraphics[width=\linewidth]{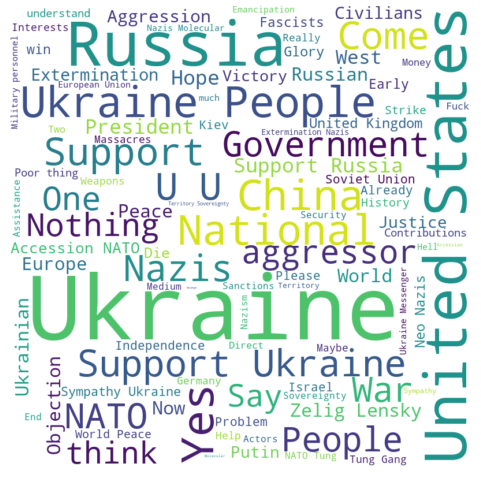}
    \caption{Tag cloud for Pro-Ukraine Weibo posts.}
    \label{fig:Figure 8}
\end{figure}

Similarly, we investigate the frequent words among Ukraine supporters as shown in Figure~\ref{fig:Figure 8}, such as \textit{Invader}, \textit{Ukrainian people}, \textit{donation}, and so on. Many Ukraine supporters hold the view that the Russia-Ukraine Conflict is an invasion war and that the Ukrainian people should fight against invaders together like what the Chinese people did 80 years ago. They also believe that it is wise for Ukraine to join NATO to obtain protection outside.

Frequent words like \textit{Gaza}, \textit{Jews}, \textit{Hamas}, and \textit{Justice} indicate the rationale of Palestine supporters. Interestingly, pro-Palestine netizens believe that Israel is a Nazi state, similar to how supporters of Russia view Ukraine. Also, they present their discontent with the United States for supporting Israel. As for people who are pro-Israel, \textit{Hamas}, \textit{terrorists}, and \textit{support} are shown frequently in their comments. Upon further querying and studying, we find that Palestine is thought to be made up of many terrorists and has to be eliminated for good among Israeli supporters.

\begin{figure}[h]
    \centering
    \includegraphics[width=\linewidth]{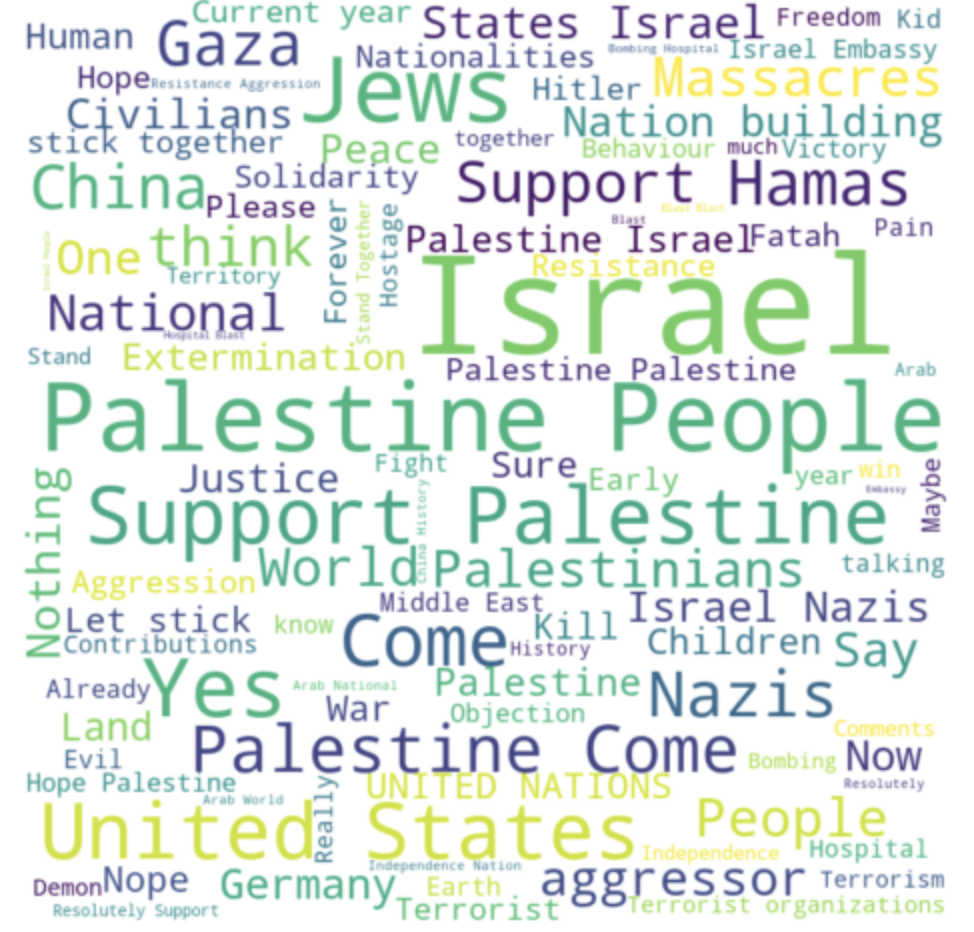}
    \caption{Tag Cloud for Pro-Palestine Weibo posts.}
    \label{fig:Figure 9}
\end{figure}

\begin{figure}[h]
    \centering
    \includegraphics[width=\linewidth]{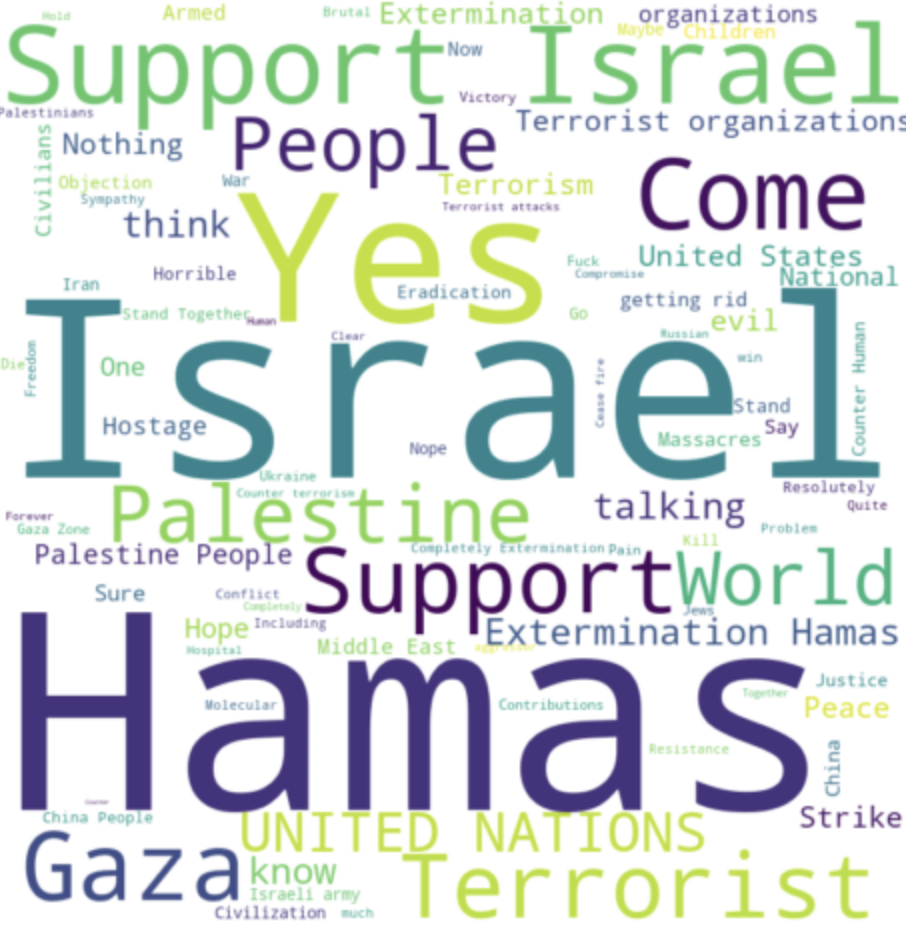}
    \caption{Tag Cloud for Pro-Israel Weibo posts. }
    \label{fig:Figure 10}
\end{figure}

\subsection{Our Findings}

The correlation between the stances of Chinese netizens on the Russia-Ukraine and Israel-Palestine conflicts reveals significant insights into the formation of public opinion in China. This suggests that opinions on one conflict may be influenced by the stances on the other. Notably, there is a marked preference for Palestine over Israel in the Israel-Palestine conflict, contrasting with the more evenly divided opinions regarding the Russia-Ukraine conflict, where a slight margin leans towards neutrality. This variation in opinion reflects the complexities involved in forming definitive views on such controversial international issues. A considerable number of users demonstrate neutrality towards both conflicts, indicating either a preference for non-partisanship or an acknowledgment of the nature of these issues.

\subsubsection{Neutrality Prevalence}

The largest individual group holds neutral opinions toward both conflicts (RU0-PI0). This suggests a significant portion of the Chinese user base either prefers to see a non-fight and peaceful situation or finds both situations too complex for a clear stance. High neutrality in individual conflicts also indicates a tendency among users to avoid taking a definitive position, which could be due to various reasons like lack of objective information, the perceived complexity of the issues, or a desire to remain impartial.

\subsubsection{Alignment in Stances}

There are notable groups where the Chinese users' stances on one conflict align with their stances on the other (\textit{e.g.}, RU1-PI1, RU2-PI2). This alignment might indicate a broader ideological or political leaning influencing their views on international conflicts~\cite{voeten2021ideology}. However, these aligned groups are not the majority, suggesting that many users view each conflict through a different lens, possibly influenced by different factors such as social media portrayal, personal experiences, or cultural background.

\subsubsection{Disparity in Support}
Although a large portion of users maintain a neutral stance on both conflicts, there is a significant inclination toward Palestine in the Palestine-Israel conflict and balance on the Russia-Ukraine conflict. This pattern suggests that while neutrality is predominant when users do take a stance, it tends to lean towards Palestine. This could reflect global sentiment trends or be influenced by regional biases. The Russia-Ukraine conflict shows a more balanced division of opinions, yet with a slight leaning towards neutrality.

\subsubsection{Media Influence}

In the previous section, we find two noticeable differences. Among the Weibo users, there is a significant inclination towards Palestine over Israel, while the support is more evenly distributed between pro-Russia and pro-Ukraine stances camps. Additionally, Weibo users demonstrate a stronger preference for Palestine and Russia when compared with the user base of the Western social platforms like Twitter and Reddit~\cite{krivivcic2023analyzing, imtiaz2022taking}.

Therefore, we aim to investigate whether differing media portrayals of the two conflicts — the Russia-Ukraine conflict and the Israel-Palestine conflict — have influenced public opinion. To this end, we have collected and analyzed news posts from five official Chinese media outlets: People's Daily, China Daily, Xinhua News Agency, CCTV News, and China Newsweek on Weibo. Our dataset, collected from Weibo posts during 2022-2023, includes 2,008 pieces covering the Russia-Ukraine conflict and 629 pieces focused on the Israel-Palestine conflict. Next, the top 25 frequent words are retrieved separately from these two datasets (shown in Table~\ref{tab:word_frequency}) to provide a foundational understanding of how each conflict is represented in the Chinese official media, the aspects that are emphasized, and the potential influence that might have on public perception.

First, there are some common frequent words in both datasets, such as ``United States'', ``Conflict'', and ``International'', indicating that Chinese media outlets may be focusing on the international reaction and the role of the United States in these two conflicts. Second, Words like ``Military'', ``Security'', ``Conflict'', ``NATO'', and ``President'' are prominent in the Russia-Ukraine conflict reports, pointing towards a focus on military actions, security issues, political figures, and international alliances. For the Palestine-Israel conflict, words like ``Conflict'', ``Hamas'', ``Death'', ``Hospital'' and ``Attack'' suggest an emphasis on the confrontational and humanitarian aspects of the conflict. Thirdly, More negative sentimental words used as ``Death'', ``Attack'', and ``Air Attack'' in the Palestine-Israel conflict reporting, might suggest a more violent vibe for the Palestine-Israel conflict. Also, this may explain why Palestine has overwhelming support on the Chinese social media platform since media outlets have been reporting huge losses in the war.

\begin{table}[ht]
\centering
\caption{Word frequency in official Chinese news reports on the two conflicts.}
\label{tab:word_frequency}
\adjustbox{max width=\linewidth}{
\begin{tabular}{lclc}
\toprule[1.1pt]
\multicolumn{2}{c}{\textbf{Report on Russia-Ukraine Conflict}} & \multicolumn{2}{c}{\textbf{Report on Palestine-Israel Conflict}} \\ 
\textbf{Word} & \textbf{Frequency} & \textbf{Word} & \textbf{Frequency} \\ \midrule
Ukraine & 2285 & Israel & 1008 \\ 
Russia & 2032 & Gaza & 939 \\ 
United States & 1431 & Strip & 728 \\ 
Show & 1114 & \textbf{Conflict} & \textbf{448} \\
Time & 941 & Palestine & 418 \\ 
China & 773 & Hamas & 345 \\ 
China Central Television & 685 & Show & 334 \\ 
President & 676 & Israeli army & 324 \\ 
Video & 604 & \textbf{Death} & \textbf{323} \\ 
Nation & 567 & Time & 254 \\ 
Claim& 561 & China& 234 \\ 
Journalist& 529 & Claim& 213 \\ 
News& 514 & United States& 211 \\ 
Military& 491 & China Central Television& 197 \\ 
Report& 487 & Hospital& 195 \\ 
Security& 458 & Say& 193 \\ 
Issue& 411 & Cause& 184 \\ 
\textbf{Conflict}& \textbf{366} & News& 178 \\ 
International& 353 & \textbf{Attack}& \textbf{176} \\ 
Provide& 350 & International& 170 \\ 
On-going& 340 & Journalist& 167 \\ 
NATO& 333 & \textbf{Air Attack}& \textbf{167} \\ 
Weibo& 331 & Civilization& 184 \\ 
Say& 317 & Amount& 159 \\ 
Putin& 307 & So far& 155 \\ \bottomrule[1.1pt]
\end{tabular}}
\end{table}

In the previous analysis of word clouds, we discuss how supporters of Russia view the Russia-Ukraine conflict as a consequence of NATO expansion, which they believe poses a threat to Russia's security. Additionally, many Chinese official news outlets have extensively reported on this viewpoint. For instance, as reported by Xinhua News Agency, a major official Chinese media outlet~\cite{xinhuanews2022ukraine},  the root cause of the Ukraine crisis is that NATO's eastward expansion threatens Russia's security. In contrast, many Western mainstream news media attribute the cause of the invasion to President Putin's personal ambitions. Some research suggests that Western media may have misrepresented Russian security interests. These studies indicate that Russia is often portrayed as motivated primarily by President Putin's contempt for democracy and an ambition to reconstruct the former Soviet Union, as discussed by \citet{mearsheimer2022causes}. In addition, \citet{zollmann2024war} thinks the Western mainstream news media has downplayed Russia’s security concerns and NATO’s militarization of the region, which has affected the portrayal of the war in Ukraine and the discussion of potential solutions. These media inclinations might significantly impact people's decisions, leading to a predominantly pro-Ukraine stance among Reddit users.

On the contrary, this observed pattern of opinions among Chinese netizens is likely influenced by broader factors, particularly media coverage and political dynamics. The way conflicts are portrayed in Chinese media, which is often influenced by the political affiliations and diplomatic stances of the Chinese government, plays a crucial role in shaping public perception. This underscores the significant impact that state-controlled media and government narratives have in framing international events and influencing public opinion. Understanding this interplay between media representation and public opinion is key to comprehensively analyzing how views are formed and swayed, especially in a context where media and state interests are closely intertwined.

It is crucial to acknowledge that there is a risk of misinterpretation or selective citation of the findings, which could aggravate existing societal or political tensions. Emphasizing one stance over another, for example, might contribute to divisive narratives and further polarize society or groups. Therefore, it is imperative to interpret the results impartially and consider the \textit{full context} of the data.

\section{Discussions and Conclusions}

This research offers an insightful analysis of Chinese netizens' views on the Russia-Ukraine and Israel-Palestine conflicts, mainly utilizing data from Weibo. The study reveals a range of opinions, with many showing a neutral stance towards both conflicts. This neutrality underscores an understanding and consideration of international disputes, potentially influenced by various factors such as media representation, political affiliations, and cultural elements. Interestingly, there is a slight but noticeable difference in preferences between the two conflicts: the Russia-Ukraine conflict sees a relatively balanced opinion with a small leaning towards neutrality, whereas in the Israel-Palestine conflict, there is a clear tendency to favor Palestine over Israel. This divergence is likely reflective of the distinct historical, political, and cultural differences inherent in each conflict.

The research also uncovers a statistical correlation between the stances on these two conflicts, suggesting that viewpoints on one issue may be interlinked or influenced by opinions on the other. The influence of media in shaping these opinions is evident, with the portrayal of these conflicts in Chinese media, likely guided by state policies and global politics, significantly impacting netizens' perspectives. 
Methodologically, the study showcases the effectiveness of advanced data mining and machine learning techniques in analyzing public opinion. This approach, adaptable to other contexts, paves the way for similar analyses globally, offering a valuable tool for understanding sentiments in the digital age.

While this study provides valuable insights into Chinese netizens' opinions on the Russia-Ukraine and Israel-Palestine conflicts, it also has several limitations. First, the data is sourced exclusively from Weibo. This presents a limitation in terms of demographic and opinion diversity. While the data may not represent the entire Chinese population, Weibo, as one of the largest social platforms in China, has a diverse user base. This wide user base can still provide a valuable snapshot of public opinion among a significant segment of the population, especially in urban and tech-savvy demographics. Second, although the study's findings are contextual to the specific time frame during which the data was collected, this temporal snapshot is valuable for understanding public opinion, especially during periods of heightened interest in the conflicts, thus providing meaningful insights into public opinion during key historical moments.

Looking forward, there are several directions for future research. Expanding the scope to include different countries and social media platforms would offer a more comprehensive view of global opinions on these conflicts. Continuous analysis is necessary to capture the evolving nature of public sentiments, especially in response to major developments in these conflicts. A comparative study of media portrayals across different countries and their impact on public opinion would provide deeper insights into the relationship between media narratives and public sentiment. Incorporating newer technologies like AI-driven sentiment analysis and natural language processing could enhance the depth, breadth, and accuracy of such studies.

Additionally, understanding these trends in public opinion can guide policymakers and diplomats in developing more effective foreign policies and communication strategies. Integrating insights from political science, media studies, and cultural studies with data science techniques would offer a more rounded understanding of public opinion formation in the digital era. In summary, this study not only clarified the perspectives of Chinese netizens on crucial international issues but also demonstrated the integral role of combining data science with social science in resolving complex public sentiments globally.

\bibliography{War_Stance_report}

\input{appendix}

\end{document}

%% file: appendix.tex
\appendix

\section{Appendix}

\subsection{Potential Broader Impact and Ethical Considerations}
There is a risk that the study could be misused by state or non-state actors for propaganda purposes. They could manipulate the data to support specific political agendas or to influence public opinion in a biased manner, especially in the context of international relations and nationalistic sentiments. As we have previously discussed, the results of this study should not be interpreted impartially and the full context of the data should be considered.

%% file: War_Stance_Detection.bbl
\begin{thebibliography}{23}
\providecommand{\natexlab}[1]{#1}
\providecommand{\url}[1]{\texttt{#1}}
\providecommand{\urlprefix}{URL }
\expandafter\ifx\csname urlstyle\endcsname\relax
  \providecommand{\doi}[1]{doi:\discretionary{}{}{}#1}\else
  \providecommand{\doi}{doi:\discretionary{}{}{}\begingroup \urlstyle{rm}\Url}\fi

\bibitem[{Ahamed et~al.(2022)Ahamed, Shakil, Lyu, Zhang, and Luo}]{ahamed2022doctors}
Ahamed, S. H.~R.; Shakil, S.; Lyu, H.; Zhang, X.; and Luo, J. 2022.
\newblock Doctors vs. Nurses: Understanding the Great Divide in Vaccine Hesitancy among Healthcare Workers.
\newblock In \emph{2022 IEEE International Conference on Big Data (Big Data)}, 5865--5870. IEEE.

\bibitem[{AlDayel and Magdy(2021)}]{aldayel2021stance}
AlDayel, A.; and Magdy, W. 2021.
\newblock Stance detection on social media: State of the art and trends.
\newblock \emph{Information Processing \& Management} 58(4): 102597.

\bibitem[{Chen et~al.(2022)Chen, Wang, Zhang, Chen, Sun, Wang, and Wang}]{chen2022public}
Chen, B.; Wang, X.; Zhang, W.; Chen, T.; Sun, C.; Wang, Z.; and Wang, F.-Y. 2022.
\newblock Public opinion dynamics in cyberspace on Russia--Ukraine War: a case analysis with Chinese Weibo.
\newblock \emph{IEEE Transactions on Computational Social Systems} 9(3): 948--958.

\bibitem[{Dalton(2013)}]{dalton2013citizen}
Dalton, R.~J. 2013.
\newblock \emph{Citizen politics: Public opinion and political parties in advanced industrial democracies}.
\newblock Cq Press.

\bibitem[{Darwish et~al.(2018)Darwish, Magdy, Rahimi, Baldwin, Abokhodair et~al.}]{darwish2018predicting}
Darwish, K.; Magdy, W.; Rahimi, A.; Baldwin, T.; Abokhodair, N.; et~al. 2018.
\newblock Predicting online islamophobic behavior after\# ParisAttacks.
\newblock \emph{The Journal of Web Science} 4(3): 34--52.

\bibitem[{Devlin et~al.(2018)Devlin, Chang, Lee, and Toutanova}]{devlin2018bert}
Devlin, J.; Chang, M.-W.; Lee, K.; and Toutanova, K. 2018.
\newblock Bert: Pre-training of deep bidirectional transformers for language understanding.
\newblock \emph{arXiv preprint arXiv:1810.04805} .

\bibitem[{Feldman(1988)}]{feldman1988structure}
Feldman, S. 1988.
\newblock Structure and consistency in public opinion: The role of core beliefs and values.
\newblock \emph{American Journal of political science} 416--440.

\bibitem[{Ghosh et~al.(2019)Ghosh, Singhania, Singh, Rudra, and Ghosh}]{ghosh2019stance}
Ghosh, S.; Singhania, P.; Singh, S.; Rudra, K.; and Ghosh, S. 2019.
\newblock Stance detection in web and social media: a comparative study.
\newblock In \emph{Experimental IR Meets Multilinguality, Multimodality, and Interaction: 10th International Conference of the CLEF Association, CLEF 2019, Lugano, Switzerland, September 9--12, 2019, Proceedings 10}, 75--87. Springer.

\bibitem[{Guan and Chen(2023)}]{guan2023moving}
Guan, T.; and Chen, X. 2023.
\newblock Moving Away from the “Repression-Resistance” Paradigm: The Effects of Civil/Uncivil Disagreements on Political Deliberation in China.
\newblock \emph{Javnost-The Public} 1--27.

\bibitem[{He and Choi(2021)}]{he2021stem}
He, H.; and Choi, J.~D. 2021.
\newblock The stem cell hypothesis: Dilemma behind multi-task learning with transformer encoders.
\newblock \emph{arXiv preprint arXiv:2109.06939} .

\bibitem[{Huang et~al.(2022)Huang, Wang, Yang, Xu et~al.}]{huang2022stance}
Huang, W.; Wang, Y.; Yang, J.; Xu, Y.; et~al. 2022.
\newblock Stance Detection Based on User Feature Fusion.
\newblock \emph{Computational Intelligence and Neuroscience} 2022.

\bibitem[{Imtiaz et~al.(2022)Imtiaz, Khan, Lyu, and Luo}]{imtiaz2022taking}
Imtiaz, A.; Khan, D.; Lyu, H.; and Luo, J. 2022.
\newblock Taking sides: Public Opinion over the Israel-Palestine Conflict in 2021.
\newblock \emph{arXiv preprint arXiv:2201.05961} .

\bibitem[{Kizilova(2022)}]{kizilova2022assessing}
Kizilova, K. 2022.
\newblock Assessing Russian public opinion on the Ukraine War.
\newblock \emph{Russian Analytical Digest} (281): 2--5.

\bibitem[{Krivi{\v{c}}i{\'c} and Martin{\v{c}}i{\'c}-Ip{\v{s}}i{\'c}(2023)}]{krivivcic2023analyzing}
Krivi{\v{c}}i{\'c}, A.; and Martin{\v{c}}i{\'c}-Ip{\v{s}}i{\'c}, S. 2023.
\newblock Analyzing Sentiment of Reddit Posts for the Russia-Ukraine War.
\newblock In \emph{2023 46th MIPRO ICT and Electronics Convention (MIPRO)}, 1709--1714. IEEE.

\bibitem[{K{\"u}{\c{c}}{\"u}k and Can(2020)}]{kuccuk2020stance}
K{\"u}{\c{c}}{\"u}k, D.; and Can, F. 2020.
\newblock Stance detection: A survey.
\newblock \emph{ACM Computing Surveys (CSUR)} 53(1): 1--37.

\bibitem[{Liu et~al.(2019)Liu, Ott, Goyal, Du, Joshi, Chen, Levy, Lewis, Zettlemoyer, and Stoyanov}]{liu2019roberta}
Liu, Y.; Ott, M.; Goyal, N.; Du, J.; Joshi, M.; Chen, D.; Levy, O.; Lewis, M.; Zettlemoyer, L.; and Stoyanov, V. 2019.
\newblock Roberta: A robustly optimized bert pretraining approach.
\newblock \emph{arXiv preprint arXiv:1907.11692} .

\bibitem[{Lyu et~al.(2022)Lyu, Wang, Wu, Duong, Zhang, Dye, and Luo}]{lyu2022social}
Lyu, H.; Wang, J.; Wu, W.; Duong, V.; Zhang, X.; Dye, T.~D.; and Luo, J. 2022.
\newblock Social media study of public opinions on potential COVID-19 vaccines: informing dissent, disparities, and dissemination.
\newblock \emph{Intelligent medicine} 2(1): 1--12.

\bibitem[{Mearsheimer(2022)}]{mearsheimer2022causes}
Mearsheimer, J.~J. 2022.
\newblock The causes and consequences of the Ukraine crisis.
\newblock \emph{The National Interest} 23.

\bibitem[{{The Moscow Times}(2022)}]{moscowtimes2022}
{The Moscow Times}. 2022.
\newblock Russia Hopes to Cleanse Ukraine of 'Nazis', Says Kremlin Spokesman.
\newblock \urlprefix\url{https://www.themoscowtimes.com/2022/02/24/russia-hopes-to-cleanse-ukraine-of-nazis-says-kremlin-spokesman-a76568}.
\newblock Accessed: 2024-01-13.

\bibitem[{Voeten(2021)}]{voeten2021ideology}
Voeten, E. 2021.
\newblock \emph{Ideology and international institutions}.
\newblock Princeton University Press.

\bibitem[{{Xinhua News Agency}(2022)}]{xinhuanews2022ukraine}
{Xinhua News Agency}. 2022.
\newblock The root cause of the Ukraine Crisis: NATO's eastward expansion threatens Russia's security \urlprefix\url{http://www.news.cn/mil/2022-02/28/c_1211590494.htm}.
\newblock Accessed on: [your access date here].

\bibitem[{Zhou et~al.(2022)Zhou, Liu, Lyu, and Luo}]{zhou2022fine}
Zhou, E.; Liu, Y.; Lyu, H.; and Luo, J. 2022.
\newblock A Fine-Grained Analysis of Public Opinion toward Chinese Technology Companies on Reddit.
\newblock \emph{arXiv preprint arXiv:2201.05538} .

\bibitem[{Zollmann(2024)}]{zollmann2024war}
Zollmann, F. 2024.
\newblock A war foretold: How Western mainstream news media omitted NATO eastward expansion as a contributing factor to Russia’s 2022 invasion of the Ukraine.
\newblock \emph{Media, War \& Conflict} 17506352231216908.

\end{thebibliography}
